\documentstyle[epsfig]{aipproc}   

\newcommand{\btau}{\mbox{\boldmath{$\tau$}}}
\newcommand{\bxi}{\mbox{\boldmath{$\xi$}}}
\newcommand{\bS}{\mbox{\boldmath{$S$}}}
\newcommand{\bz}{\mbox{\boldmath{$\zeta$}}}
\newcommand{\bJ}{\mbox{\boldmath{$J$}}}
\newcommand{\JO}{\mbox{\boldmath{$J^{0}$}}}
\newcommand{\cD}{{\cal D}}
\newcommand{\cJ}{{\cal J}}

\begin{document}

\title{Public key cryptography and error correcting codes as Ising models}
\author{ David~Saad$^{*}$, Yoshiyuki~Kabashima$^{\dagger}$ and Tatsuto
Murayama$^{\dagger}$} \address{ $^{*}$The Neural Computing Research Group,
Aston University, Birmingham B4 7ET, UK. \\ $^{\dagger}$Dept.~of
Comp.~Intel.~\& Syst.~Sci., Tokyo Institute of
Technology, Yokohama 2268502, Japan.  } 
\maketitle
\begin{abstract}
We employ the methods of statistical physics to study the performance
of Gallager type error-correcting codes. In this approach, the
transmitted codeword comprises Boolean sums of the original message
bits selected by two randomly-constructed sparse matrices. We show
that a broad range of these codes potentially saturate Shannon's bound
but are limited due to the decoding dynamics used. Other codes show
sub-optimal performance but are not restricted by the decoding
dynamics.  We show how these codes may also be employed as a practical
public-key cryptosystem and are of competitive performance to modern
cyptographical methods.
\end{abstract}
%
%
Error-correcting codes are of significant practical importance as they
provide mechanisms for retrieving the original message after possible
corruption due to noise during transmission. They are being used
extensively in most means of information transmission from satellite
communication to the storage of information on hardware devices. The
coding efficiency, measured in the fraction of informative
transmitted bits, plays a crucial role in determining the speed of
communication channels and the effective storage space on hard-disks.
Rigorous bounds~\cite{Shannon} have been derived for the maximal
channel capacity for which codes, capable of achieving arbitrarily
small error probability, can be found. However, most existing
practical error-correcting codes are significantly far from saturating
this bound and the quest for more efficient error-correcting codes has
been going on ever since.

One family of codes, introduced originally by
Gallager~\cite{Gallager}, and abandoned in favour of other codes due
to the limited computing facilities of the time, has recently been
re-introduced~\cite{MacKay}, showing excellent performance with
respect to most existing codes.  In fact, it has recently been
discovered that irregular constructions of Gallager's code result in
better performance than any other method~\cite{Richardson,casdec} and
nearly saturate Shannon's bound for infinite message size.
Gallager-type methods are generally based on the introduction of
random sparse matrices for generating the transmitted codeword as well
as for decoding the received corrupted codeword. Various decoding
methods have been successfully employed; we will mainly focus here on
the leading technique of Belief Propagation (BP)~\cite{pearl}.

Most studies of Gallager-type codes conducted so far have been carried
out via numerical simulations. Some analytical results have been
obtained via methods of information theory~\cite{MacKay}, setting
bounds on the performance of certain code types, and by
combinatorial/statistical methods~\cite{Richardson}. Here we analyze
their typical performance for several parameter choices via the
methods of statistical physics, and validate the analytical solutions
against results obtained by the Thouless-Anderson-Palmer (TAP)
approach~\cite{TAP} to diluted systems and via numerical methods.

In a general scenario, the $N$ dimensional Boolean message $\bxi$ is
encoded to the $M$ dimensional vector $\JO$ which is then transmitted
through a noisy channel with flipping probability $p$ per bit (other
noise types may also be considered). The received message $\bJ$ is
decoded to retrieve the original message.

One can identify several slightly different versions of Gallager-type
codes. The one used here, termed the MN code\cite{MacKay} is based on
choosing two randomly-selected sparse matrices $A$ and $B$ of
dimensionality $M\!\times \!N$ and $M\!\times\! M$ respectively; these
are characterized by $K$ and $L$ non-zero unit elements per row and
$C$ and $L$ per column respectively. The finite, usually small,
numbers $K$, $C$ and $L$ define a particular code; both matrices are
known to both sender and receiver. Encoding is carried out by
constructing the modulo 2 inverse of $B$ and the matrix $B^{-1}A$
(modulo 2); the vector $\JO \! =\! B^{-1}A \ \bxi$ (modulo 2, $\bxi$
in a Boolean representation) constitutes the codeword. Decoding is
carried out by taking the product of the matrix $B$ and the received
message $\bJ\! = \!  \JO \! +\! \bz$ (modulo 2). The equation
\begin{equation}
\label{eq:decoding}
A\bxi + B\bz = A\bS + B\btau  \  \  \mbox{(mod 2)}, \
\end{equation}
is solved via the iterative methods of BP~\cite{MacKay} to obtain the
most probable Boolean vectors $\bS$ and $\btau$; BP methods in this
context have recently been shown to be identical to a TAP based
solution of a similar physical system~\cite{us_sourlas}.

The similarity between error-correcting codes of this type and Ising
spin systems was first pointed out by Sourlas\cite{Sourlas}, who
formulated the mapping of a simpler code onto an Ising spin system
Hamiltonian.  To facilitate the current investigation we first map the
problem onto that of an Ising model with finite connectivity. We
employ the binary representation $(\pm1)$ of the dynamical variables
$\bS$ and $\btau$ and of the vectors $\bJ$ and $\JO$ rather than the
Boolean $(0,1)$ one; the vector $\JO$ is generated by taking products
of the relevant binary message bits $J^{0}_{\mu} \! = \!  \prod_{i\in
\mu} \xi_{i} $, which correspond to the non-zero elements of $B^{-1}A$,
producing a binary version of $\JO$. As we use statistical mechanics
techniques, we consider the message and codeword dimensionality ($N$
and $M$ respectively) to be infinite, keeping the ratio $R \!=\!
N/M$, which constitutes the code rate, finite. Using the thermodynamic
limit is natural as Gallager-type codes are used to transmit long
($10^{4}\!-\!10^{5}$) messages, where finite size corrections are
likely to be negligible.  We examine the Hamiltonian
\begin{equation}
\label{eq:Hamiltonian}
{\cal H} = \sum_{\mu,\sigma}
      \; \cD_{\mu \sigma} \ \delta
      \biggl[-1 \ ; \ \cJ_{\mu \sigma} \nonumber \prod_{i\in\mu} S_{i}
      \prod_{j\in\sigma} \tau_{j}
      \biggr] - \frac{F_s}{\beta} \sum_{i=1}^{N} S_i -
      \frac{F_{\tau}}{\beta} \sum_{j=1}^{M} \tau_j \ .
\end{equation}
The tensor product $\cD_{\mu\sigma} \cJ_{\mu\sigma}$, where
$\cJ_{\mu\sigma} \! = \!  \prod_{i\in\mu}{\xi_{i}}\prod_{j\in\sigma}
\zeta_{j}$ and $\sigma=\langle j_{1},\ldots j_{L} \rangle $, is the
binary equivalent of $A\bxi \!  + \!  B\bz$, treating both signal
($\bS$ and index $i$) and noise ($\btau$ and index $j$)
simultaneously. Elements of the sparse connectivity tensor $\cD_{\mu
\sigma}$ take the value 1 if the corresponding indices of both signal
and noise are chosen (i.e., all corresponding indices of $A$ and $B$
are 1) and 0 otherwise; it has $C$ unit elements per $i$-index and $L$
per $j$-index, representing the system's degree of connectivity.  The
$\delta$ function provides $1$ if the selected sites' product
$\prod_{i\in\mu} S_{i} \prod_{j\in\sigma}\tau_{j}$ is in disagreement
with $\cJ_{\mu\sigma}$, recording an error, and $0$ otherwise. Note
that this term is not frustrated, as there are $M\!  +\!N$ degrees of
freedom and $M$ constraints (\ref{eq:decoding}). The two terms on the
right represent our prior knowledge in the case of biased messages
$F_s$ and of the noise level $F_{\tau}$, and require assigning certain
values to these additive fields. The choice of $\beta\!  \rightarrow
\!  \infty$ imposes the restriction of Eq.(\ref{eq:decoding}), while
the last two terms remain finite. Note that the noise dynamical
variables $\btau$ are irrelevant to measuring the retrieval success $
m = \frac{1}{N} \ \left\langle \sum_{i=1}^{N} \ \xi_{i} \ \mbox{sign}
\left\langle S_{i} \right\rangle_{\beta} \right\rangle_{\xi} \ .$ The
latter monitors the normalized mean overlap between the Bayes-optimal
retrieved message, corresponding to the alignment of $\left\langle
S_{i} \right\rangle_{\beta}$ to the nearest binary
value\cite{Sourlas}, and the original message; the subscript $\beta$
denotes thermal averaging.  The selection of elements in $\cD$
introduces disorder to the system; we calculate the partition function
${\cal Z} ({\cD},\mbox{\boldmath $J$}) = \mbox{Tr}_{\{\bS,\btau\}}
\exp [-\beta {\cal H}]$ averaged over $\cD$, $\bxi$ and $\bz$ using
the replica method~\cite{us_sourlas}. We employ the replica symmetry
ansatz to obtain a set of saddle point equations with respect to the
emerging continuous order parameters, representing local field
probability distributions and the respective conjugate
distributions~\cite{us_gallager}.

For unbiased messages and either $K\!\ge\! 3$ $(L \! \ge \! 2)$ or
$L\!\ge\! 3$ $(K \! \ge \! 2)$ we obtain both the ferromagnetic and
paramagnetic solutions either by applying the TAP approach or by
solving the saddle point equations numerically. The former was carried
out at the values of $F_{\tau}$ and $F_s\!=\! 0$) which correspond to
the true noise and input bias levels (for unbiased messages $F_s\!=\!
0$) and thus to Nishimori's condition\cite{Nishimori}.  This is
equivalent to having the correct prior within the Bayesian
framework~\cite{Sourlas}.

The most interesting quantity to examine is the maximal code rate, for
a given corruption process, for which messages can be perfectly
retrieved.  This is defined in the case of $K,L\!\ge\! 3$ by the value
of $R \! = \! K/C \! = \! N/M$ for which the free energy of the
ferromagnetic solution becomes smaller than that of the paramagnetic
solution, constituting a first order phase transition.  The critical
code rate obtained $ R_{c}\!=\! 1\!-\!H_{2}(p)\!=\! 1\!+\!\left(p
\log_{2} p \!+\!(1-p) \log_{2} (1-p) \right), $ coincides with {\em
Shannon's capacity}.

The MN code for $K,L \ge 3$ seems to offer optimal performance.
However, the main drawback is rooted in the co-existence of the stable
$m=1,0$ solutions, which implies that from most initial conditions the
system will converge to the undesired paramagnetic solution. Studying
the ferromagnetic solution numerically shows a highly limited basin of
attraction, which becomes smaller as $K$ and $L$ increase, while the
paramagnetic solution at $m=0$ {\em always} enjoys a wide basin of
attraction.

\begin{figure}[t]
\begin{minipage}{0.4\linewidth}
%
\begin{picture}(12,10)(0,50)
\put(148,-145){\epsfxsize=100mm  \epsfbox{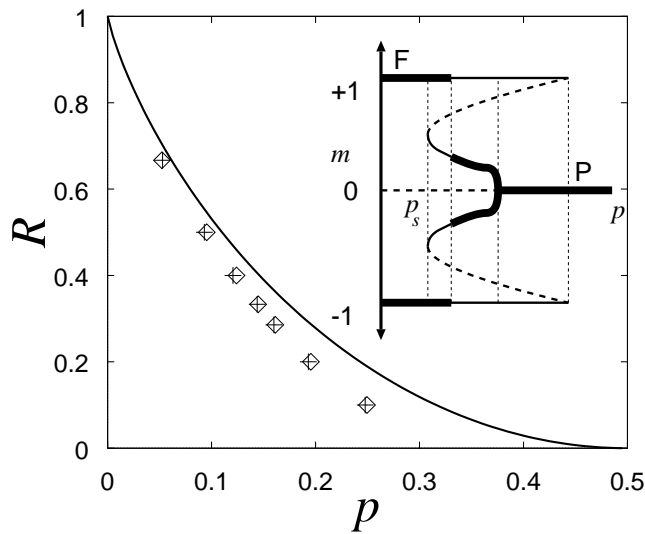}}
\put(303,-75){\epsfxsize=40mm  \epsfbox{PT_K2L2.eps}}
\end{picture}
%
\end{minipage}

\vspace*{-3mm}
\begin{minipage}{0.4\linewidth}
\parbox{0.5\linewidth}{
\caption{\footnotesize Critical transmission rate as a function of
$p$, obtained analytically ($\Diamond$) and via BP ($+$) iterative
solutions ($N\!=\!  10^4$) for unbiased messages (averaged over 10
different initial conditions); error bars are smaller than the symbol
size. Shannon's bound (solid line) is shown for reference.  Inset: The
ferromagnetic ({\sf F}) and paramagnetic ({\sf P}) solutions as
functions of $p$; thick and thin lines denote stable solutions of
lower and higher free energies respectively, dashed lines correspond
to unstable solutions. Lines between the $m\!=\!\pm 1$ and $m\!=\! 0$
axes correspond to sub-optimal ferromagnetic solutions. } }

             \end{minipage}     \vspace*{-4mm}       
\end{figure}

Studying the case of $K\!=\! L\!=\! 2$ , indicates the existence of
paramagnetic and ferromagnetic solutions depicted in the inset of
Fig.1. For corruption probabilities $p\!>\!p_{s}$ one obtains either a
dominant paramagnetic solution or a mixture of ferromagnetic
($m\!=\!\pm 1$) and paramagnetic ($m\!=\! 0$) solutions.  Reliable
decoding may only be obtained for $p\!<\!p_{s}$, which corresponds
to a spinodal point, where a unique ferromagnetic solution emerges at
$m\!=\! 1$ (plus a mirror solution at $m\!=\!  -1$). Initial
conditions for both simulations (TAP/BP) and the numerical solutions
were chosen almost randomly, with a slight bias of ${\cal
O}(10^{-12})$, in the initial magnetization. The results
obtained point to the existence of a unique pair of global solutions
to which the system converges (below $p_{s}$) from {\em all initial
conditions}.

The main question that emerges is the possibility of producing more
complicated constructions for which the spinodal point is closer to
Shannon's critical flip rate. This has been based mainly on the
introduction of irregular constructions~\cite{Richardson,casdec},
where the number of unit elements per row/column in the matrices $A$
and $B$ is not fixed. Analytical investigation~\cite{irregular} aimed
at optimising the construction are yet to provide a principled method
for carrying out the optimisation.

The study of  parity check codes and the insight gained
from the analysis led us to suggest the potential use of a similar
system as a public-key cryptosystem~\cite{crypto}.

Public-key cryptography is based on a distribution of a public key
which may be used to encrypt messages in a manner that can only be
decrypted, in practical time scales, by the service provider. Several
quite safe and efficient cryptosystems are currently in use such as
RSA, elliptic-curves and the McEliece cryptosystem ~\cite{book}, most
of which are based on number theory methods. Public-key cryptography
plays an important role in many aspects of modern information
transmission, for instance, in the areas of electronic commerce and
internet-based communication. It enables the service provider to
distribute a public key which may be used to encrypt messages in a
manner that can only be decrypted by the service provider.

In the suggested cryptosystem, a plaintext represented by an $N$
dimensional Boolean vector $\bxi \in (0,1)^N$ is encrypted to the $M$
dimensional Boolean ciphertext $\bJ$ using a predetermined Boolean
matrix $G$, of dimensionality $M\times N$, and a corrupting $M$
dimensional vector $\bz$, whose elements are 1 with probability $p$
and 0 otherwise, in the following manner $\bJ = G \ \bxi \ + \bz \ , $
where all operations are (mod 2). The matrix $G$ and the probability
$p$ constitute the public key; the corrupting vector $\bz$ is chosen
at the transmitting end. The matrix $G$, which is at the heart of the
encryption/decryption process is constructed by choosing two
randomly-selected sparse matrices $A$ ($M\!\times \!N$) and $B$
($M\!\times\! M$), and a dense matrix $D$ ($N\!\times \!N$), defining
$ G \! =\! B^{-1}A D \ \ \mbox{(mod 2)} \ . $ The matrices $A$ and $B$
are generally characterised by $K$ and $L$ non-zero unit elements per
row and $C$ and $L$ per column respectively; all other elements are
set to zero. The finite, usually small, numbers $K$, $C$ and $L$
define a particular cryptosystem.  The dense invertible Boolean matrix
$D$ is arbitrary and is added for improving the system's security. It
may be constructed as $D=T P$, where $T$ and $P$ are $N \times N$
triangular and random permutation matrices respectively, for
minimising the computational costs. All matrices are known only to the
authorised receiver.  Suitable choices of probability $p$ will depend
on the maximal achievable rate for the particular cryptosystem as
discussed below.

The authorised user may decrypt the ciphertext $\bJ$ by taking the
(mod 2) product $B \bJ \!=\! A (D\bxi)\! + \!  B\bz$, and finding the
most probable solution to Eq.(\ref{eq:decoding}) using the methods of BP;
obtaining the estimate of $\bxi$ is obtained by taking the product of
the $ (D\bxi)$ estimate and $D^{-1}$. Studying the case of
$K\!=\!L\!=\!2$ and $p<p_{s}$ we learned that iterative BP
decoding converges to the ferromagnetic solution from {\em any}
initial conditions.  Cryptosystems with other $K,L$ values generally
suffer from a decreasingly small basin of attraction as $K$ and $L$
increase, although specific matrices with higher $K$ and $L$ values
(such as in~\cite{casdec}) may still be used successfully.

The cryptosystem offers a guaranteed convergence to the plaintext
solution, in the thermodynamic limit $N\rightarrow \infty$, as long as
$p<p_{s}$.  The main consequence of finite plaintexts would be a
decrease in the allowed corruption rate. Experimental results with
systems size as small as $N\!=\!1000$ still show good performance.

The unauthorised receiver, on the other hand, faces the task of
decrypting the ciphertext $\bJ$ knowing only $G$ and $p$.  The
straightforward attempt to try all possible $\bz$ constructions is
clearly doomed, provided that $p$ is not vanishingly small, giving
rise to only a few corrupted bits. We study the
problem by exploiting the similarity between the task at hand and the
error-correcting model suggested by Sourlas\cite{Sourlas}. 
In this case, the matrix $G$ generated in the case of $K\!\!=\!\!
L\!\!=\!\!  2$ is dense and has a certain distribution of unit
elements per row.  The fraction of rows with a low (finite, not of
${\cal O} (N)$) number of unit elements vanishes as $N$ increases,
allowing one to approximate this scenario by the diluted random energy
model studied in~\cite{us_sourlas}.

To investigate the typical properties of this (frustrated) model, we
calculate again the partition function and the free energy by
averaging over the randomness in choosing the plaintext, the
corrupting vector and the choice of the random matrix $G$ (being
generated by a product of two sparse random matrices). To assess the
likelihood of obtaining spin-glass/ferromagnetic solutions, we
calculated the free energy landscape (per plaintext bit - $f$) as a
function of overlap $m$.  This can be carried out straightforwardly
using the analysis of \cite{us_gallager}, and provides the
golf-course-like energy landscape with a relatively flat area around
the one-step replica symmetry breaking (frozen) spin-glass solution
and a very deep but extremely narrow area, of ${\cal O} (1/N)$, around
the ferromagnetic solution~\cite{crypto}.

It is worthwhile mentioning that this free-energy landscape may be
related directly to the marginal posterior $P(S_{i}\!=\!1 | \bJ) \
1\!\le\! i\! \le\! N $ and is therefore indicative of the difficulties
in obtaining ferromagnetic solutions when the starting point for the
search is not infinitesimally close to the original plaintext (which
is clearly highly unlikely).  Numerical studies of similar energy
landscapes show that the time required increases exponentially with
the system size~\cite{Parisi}.

Most attacks on this cryptosystems, by an unauthorised user, will face
the same difficulty: without explicit knowledge of the current
plaintext and/or the decomposition of $G$ to the matrices $A$, $B$ and
$D$ it will require an exponentially long time to decipher a specific
ciphertext.  We investigated attacks of several types, some of which
appear in~\cite{crypto}, concluding that the suggested system is
secure.

A brief comparison of our method and the leading technique of
RSA~\cite{book} shows that: 1) RSA decryption takes ${\cal O} (N^3)$
operations while our method only requires ${\cal O} (N \log N)$
operations.  2) Encryption costs are of ${\cal O} (N^2)$ (as in RSA);
inverting the matrices $B$ and $D$ is carried out only once and is of
${\cal O} (N^{3})$.  Two drawbacks of our method: 1) The public key is a dense
matrix of dimensionality $M\!\times\!  N$. However, as public key
transmission is carried out only once we do not expect it to be of
great significance.  2) The ciphertext/plaintext bit ratio is greater
than one (as is the case in RSA).  Choosing the $N/M$ ratio is in the
hands of the user and is related to the security level required. In
addition, the increased transmission time is compensated by a very
fast decryption and the added robustness against noise.

We discussed the relation between Ising models, certain
error-correcting codes and public-key cryptosystems. Important aspects
that are yet to be investigated include the relation between our
results and the bounds obtained in the information theory literature
for error-correcting codes, finite size effects and methods for
alleviating the drawbacks of the new cryptosystem.

\vspace{3mm}
{\footnotesize 
%
\hspace*{-1.5em} Support by
JSPS-RFTF (YK), The Royal Society and EPSRC-GR/N00562 (DS) is acknowledged.}


\end{document}